\documentclass{article}

\usepackage{arxiv}

\usepackage[utf8]{inputenc} 
\usepackage[T1]{fontenc}    
\usepackage{hyperref}       
\usepackage{url}            
\usepackage{booktabs}       
\usepackage{amsfonts}       
\usepackage{nicefrac}       
\usepackage{microtype}      
\usepackage{lipsum}

\usepackage{siunitx}
\usepackage[caption=false,font=footnotesize]{subfig}
\usepackage[labelformat=parens,labelsep=quad,skip=3pt]{caption}
\usepackage{graphicx}
\usepackage{color,soul}
\usepackage{algorithm,algpseudocode}
\usepackage[noend]{algcompatible}
\usepackage{bm}
\usepackage{mathtools}
\usepackage{xspace}
\usepackage[english]{babel}

\title{Privacy-preserving Federated Learning based on Multi-key Homomorphic Encryption}

\author{
  Jing Ma \\
  Xidian University, China\\
   \And
 Si-Ahmed Naas \\
  Aalto University, Finland \\
  \And
 Stephan Sigg \\
  Aalto University, Finland \\
  \And
 Xixiang Lyu\thanks{Corresponding author. E-mail address: xxlv@mail.xidian.edu.cn }\\
  Xidian University, China\\
}

\begin{document}
\maketitle

\begin{abstract}

With the advance of machine learning and the internet of things (IoT), security and privacy have become key concerns in mobile services and networks. 
Transferring data to a central unit violates privacy as well as protection of sensitive data while increasing bandwidth demands.
Federated learning mitigates this need to transfer local data by sharing model updates only. 
However, data leakage still remains an issue. 
In this paper, we propose xMK-CKKS, a multi-key homomorphic encryption protocol to design a novel privacy-preserving federated learning scheme.
In this scheme, model updates are encrypted via an aggregated public key before sharing with a server for aggregation. 
For decryption, collaboration between all participating devices is required.
This scheme prevents privacy leakage from publicly shared information in federated learning, and is robust to collusion between $k<N-1$ participating devices and the server. 
Our experimental evaluation demonstrates that the scheme preserves model accuracy against traditional federated learning as well as secure federated learning with homomorphic encryption (MK-CKKS, Paillier) and reduces computational cost compared to Paillier based federated learning. 
The average energy consumption is 2.4 Watts, so that it is suited to IoT scenarios.
\end{abstract}

\section{Introduction}
It is estimated that the number of IoT devices worldwide will reach 75 billion by 2025~\cite{TP-toolbox-web}. 
This development will cause an explosive growth of distributed data generated by these IoT devices. 

This setting holds enormous potential to enable novel applications in many different domains, including smart healthcare, smart home, image classification, language representation, or traffic accident detection. 
The ubiquitous distributed instrumentation of objects and environments with sensing and computational capabilities fosters smart IoT applications combined with distributed learning technologies.
Exploiting such massive data for activity recognition and situational awareness has been an active field of research in recent years~\cite{bulling2014tutorial}. 
 
\begin{figure}
    \subfloat[Federated learning without encryption]{\includegraphics[width=.31\textwidth]{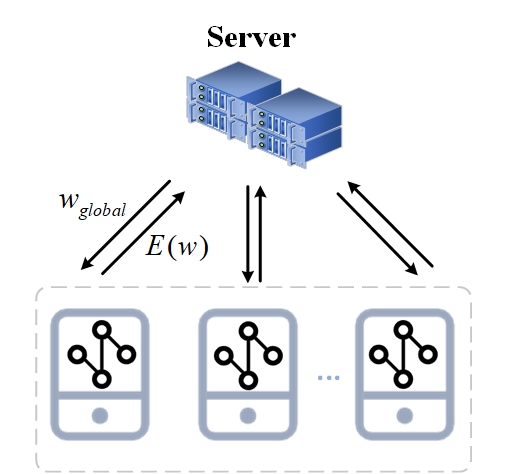}\label{fig:Comparisona}}
    \hfill
    \subfloat[Homomorphic encryption (HE)-based federated  learning using the same keys for all parameters]{\includegraphics[width=.31\textwidth]{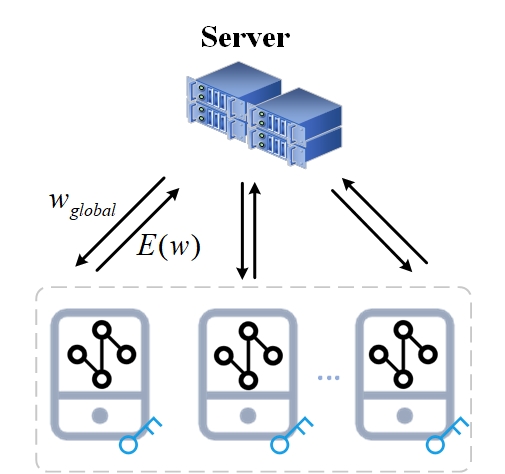}\label{fig:Comparisonb}}
    \hfill
    \subfloat[Multi-key HE-based federated learning using distinct encryption keys. Decryption is possible only combining all keys]{\includegraphics[width=.31\textwidth]{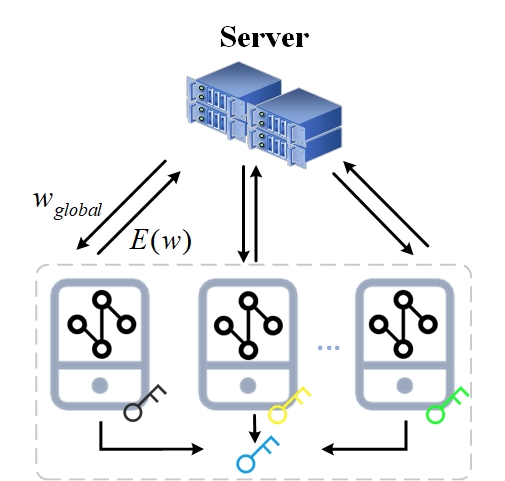}\label{fig:Comparisonc}}
    \caption{Different federated learning schemes with various types of encryption. We propose multi-key homomorphic encryption in which parameters are not able to decrypt other parameters encrypted weights.}
    \label{fig:Comparison}
\end{figure}

However, IoT data often contains private information such as medical records and personal activities, which might be leaked through distributed machine learning. 
To make use of large-scale distributed data in such scenario, federated learning~\cite{konevcny2016federated} enhances privacy protection by avoiding to share local data during the training process (Fig~\ref{fig:Comparisona}).
In particular, not the data, but model updates (weights or gradients) which are calculated locally at distributed devices, are then shared with a server to update a global model. 
Nonetheless, private information of the local data may still leak from the model updates~\cite{melis2019exploiting,zhu2019deep,hitaj2017deep}. 
Any participating device able to observe and analyze these updates may thus pose a threat to the privacy protection interests of other devices, which ultimately discourages participation in the training of a distributed model.

Recent attempts to address this issue by applying cryptographic techniques are not suitable for distributed mobile IoT scenarios, where storage space and processing capabilities of distributed IoT devices are constrained. 
While homomorphic encryption (HE) enables the computation on encrypted model updates~\cite{aono2017privacy,zhang2020privacy}, all participants in these works share the same public key for encryption and, more importantly, the same secret key for decryption (Fig~\ref{fig:Comparisonb}). In this way, the encrypted data of the participant can be decrypted by any other participant.
Therefore, traditional homomorphic encryption schemes in federated learning can not resist attacks from internal curious devices as well as collusion attacks between devices and the server.

Unlike the traditional homomorphic encryption, multi-key homomorphic encryption (MK-HE) allows different participants to use different keys for encryption. The decryption of the aggregated ciphertext requires the collaboration of all participants, thus provides higher security. MK-CKKS is a state-of-the-art multi-key fully homomorphic encryption scheme which supports approximate fixed-point arithmetic.  
We found that MK-CKKS has the risk of privacy leakage when it is used in the federated learning scenario to protect privacy of model updates. In this paper, to solve this problem, we propose xMK-CKKS, which improves the state-of-the-art multi-key homomorphic encryption scheme MK-CKKS by requiring an aggregated public key for encryption as well as the decryption share for secure and consented decryption. The proposed xMK-CKKS provides strong security and doesn't require interactions among the participans, therefore is suitable for federated learning scenarios. 
In this way, a privacy-preserving federated learning scheme based on xMK-CKKS is proposed (Fig~\ref{fig:Comparisonc}). The model updates are homomorphically encrypted before sharing for aggregation. 
Decryption of aggregated results is only possible by all contributing parties collaboratively. 
Therefore, the scheme is robust against attacks from the participants and also against collusion attacks between participants and the server. 
It guarantees confidentiality of model updates and thereby provides strong privacy protection for federated learning compared with the traditional homomorphic encryption based federated learning and also the federated learning based on MK-CKKS.
Finally, we verify this secure federated learning scheme experimentally in a realistic IoT scenario for smart healthcare, where remote participating devices are not necessarily all interconnected. Specifically, our contributions are: 
\begin{itemize}
\item[(1)] We propose xMK-CKKS, which improves the MK-CKKS scheme by setting an aggregated public key which is the sum of individual public key for encryption. The decryption of an aggregated ciphertext requires the decryption shares which implicitly contain the information of individual secret key and the aggregated cipheretexts from all participants, thus has no threat to individual ciphertext.

\item[(2)] We propose a privacy-preserving federated learning scheme based on xMK-CKKS. 
The scheme guarantees confidentiality of model updates in the honest-but-curious setting by encrypting them with multi-key homomorphic encryption. This scheme is robust against attacks from the participants and also against collusion attacks between $k<N-1$ participating devices and the server.

\item[(3)] We evaluate the scheme on Jetson Nano IoT devices for a realistic distributed learning scenario and against state-of-the-art federated learning schemes based on homomorphic encryption. 
Results show significant reduction in computational load and reasonable energy consumption while maintaining the accuracy.
\end{itemize}

\section{Related work}
Our work is mostly related to prior research conducted in the domains of multi-key homomorphic encryption, as well as federated learning, in particular, with respect to privacy preserving federated learning schemes.  
\subsection{Multi-key homomorphic encryption}
Popular additive homomorphic encryption (HE) schemes include Goldwasser and Micali~\cite{goldwasser2019probabilistic}, Paillier~\cite{paillier1999public}, Damgard and Jurik~\cite{damgaard2001generalisation}, or Kawachi et al.~\cite{kawachi2007multi}.
The first fully homomorphic encryption (FHE) has been proposed by Gentry~\cite{gentry2009fully}, which has seen a lot of follow-up improvements to reduce computational load. 
Homomorphic encryption schemes can be symmetric (same key for encryption and decryption), as well as asymmetric (different keys)~\cite{rothblum2011homomorphic}.

Unlike traditional homomorphic encryption that evaluates the arithmetic circuits of ciphertexts encrypted with the same key, multi-key homomorphic encryption (MK-HE) allows different parties to use different keys for encryption. Decryption requires the collaboration of all participants. 
López-Al et al.~\cite{lopez2012fly} proposed the first MK-FHE scheme known as LTV12, which is based on the NTRU encryption system~\cite{hoffstein1998ntru}. 
It uses relinearization and modulus switching technologies to obtain a leveled multi-key fully homomorphic encryption scheme, but suffers from high computational complexity of decryption operations.
While more efficient protocols, such as DHS16~\cite{doroz2016homomorphic} have been proposed, NTRU-based MK-FHE schemes are not practical due to their large decryption complexity and high communication load.
The use of other types of FHE schemes for MK-FHE has been first discussed by Michael et al.~\cite{clear2015multi}. 
Further simplifications in the ciphertext extension process, and multiple rounds of multi-party computation (MPC) have been introduced in~\cite{mukherjee2016two}, however, the size of the ciphertext in these schemes increases exponentially with the number of participants, which causes communication and storage costs that are infeasible for application in a practical setting. 
In 2019, Chen et al.~\cite{chen2019efficient} proposed MK-BFV and MK-CKKS, multi-key variants of the fully homomorphic encryption schemes BFV~\cite{brakerski2012fully,fan2012somewhat} and CKKS ~\cite{cheon2017homomorphic}. 
In these schemes, the length of the ciphertext increases linearly with the number of participants. 
The implementations of MK-BFV and MK-CKKS provide the first experimental results of MK-HE with packed ciphertexts.
However, the distributed decryption of a multi-key ciphertext needs additional secure method, otherwise has the risk of privacy leakage. 
The proposed scheme xMK-CKKS in this paper provides simple and secure decryption for a aggregated ciphertext generated by only addition.

\subsection{Privacy-preserving federated learning}
Federated learning was first introduced by Kone{\v{c}}n{\`y} et al.~\cite{konevcny2016federated} in 2016, as a distributed machine learning scheme in which distributed devices collaborate with a central coordinator in sharing locally trained updates to a global model for aggregation. 
Several improvements have since been proposed. 
To reduce the uplink communication cost in federated learning, structured and sketched updates were introduced~\cite{konevcny2016federated1}. 
In addition, McMahan et al. proposed Federated Averaging (FedAvg), which achieves a reduction in the communication load by one to two orders of magnitude through iterative model averaging.  
Their implementation is robust to unbalanced and not independent and identically distributed (non-iid) data, and further obfuscates the shared information through aggregated model updates~\cite{mcmahan2017communication}.

On the other hand, concerns about information leakage and privacy issues of federated learning have been raised. 
Aono et al.~\cite{aono2017privacy} pointed out an adversary capable of observing even only a small fraction of the gradients, can lead to information leakage and data privacy issues for the distributed devices sharing their updates. 
Indeed, Melis et al.~\cite{melis2019exploiting} demonstrated that in distributed federated learning, even if only model update information is shared, sensitive information of the participant may be leaked. 
It is even possible to recover original training data from the publicly shared gradients~\cite{zhu2019deep}. 

To address this weakness and to enhance the privacy protection of federated learning, cryptographic methods have been proposed, such as multi-party computation~\cite{cramer2015secure}, differential privacy~\cite{10.1007/11787006_1,dwork2006calibrating,dwork2006our}, as well as homomorphic encryption~\cite{rivest1978data}, to protect the shared model updates.

In secure multi-party computation, a group of mutually distrustful parties $u\in\mathcal{U}$ collaborate to compute an aggregated sum $\mathcal{A}=\sum_{u\in\mathcal{U}}x_u$ for their private values $x_u$, without revealing $x_u$.  
The scheme of secure aggregation has been adapted for federated learning, for instance, in~\cite{bonawitz2016practical,bonawitz2017practical}, to ensure that globally, an individual update from any distributed party can not be derived from the aggregation of all updates. 
Related to this, Li et al.~\cite{li2020privacy} proposed to chain local gradient updates across all remote participating devices in order to mask the individual values. 
However, the protocol requires secure and reliable communication channels and honest devices. 
In addition, the greedy generation of the chain of devices might fail in partly disconnected topologies. 
Xu et al.~\cite{xu2019verifynet} proposed privacy preserving and verifiable collaborative learning, which uses double masking to ensure the confidentiality of the individual gradients. 
In particular, distributed devices secretly establish random numbers for each device pair, to mask their local data. 
However, the masking process is computationally expensive, and constraints practical use by requiring that all devices are within pairwise communication range. 
Furthermore, a trusted third party is needed to initially establish asymmetric key pairs.

Differential privacy~\cite{abadi2016deep} may protect the privacy of individual data of deep learning algorithms via a differentially private gradient descent mechanism that adds noise to gradient updates. 
Geyer et al.~\cite{geyer2017differentially} developed an implementation of differential privacy for federated learning, which realizes client differential privacy guarantees at the cost of severely reduced model performance. 
 
Dowlin et al.~\cite{gilad2016cryptonets} presented a neural network based on homomorphic encryption~\cite{lopez2012fly}.  
In particular, the model inference from a pre-trained network is computed on encrypted data by an untrusted party. 
Network training, however, has to be conducted offline and does not involve the untrusted party.
Improved schemes have been proposed in~\cite{li2017multi},~\cite{aono2017privacy} using addictive homomorphic encryption to protect model updates, by Zhang et al.~\cite{zhang2020privacy} exploiting privacy-preserving and verifiable federated learning using the Paillier cryptosystem~\cite{paillier1999public}, as well as by Froelicher et al.~\cite{froelicher2020drynx} who proposed a decentralized system for privacy-conscious statistical analysis on distributed datasets by applying the ElGamal Elliptic Curve additive homomorphic cryptosystem~\cite{elgamal1985public}. 
In these approaches, all participating devices share the same encryption and decryption key, so that private information may leak among devices. 
Furthermore, any curious party colluding with the server will breach the privacy of other parties. 
In this paper, we solve these problems by applying multi-key homomorphic encryption. Our privacy-preserving federated learning scheme prevents information leakage from other devices and even collusion between $k<N-1$ devices and the server.

\section{Preliminaries}\label{sec:preliminaries}
We propose xMK-CKKS, a multi-key homomorphic encryption scheme based on the MK-CKKS, to increase privacy protection in distributed machine learning, specifically in federated learning.    
In the following, we discuss our assumptions on the federated learning and introduce the concept of the MK-CKKS homomorphic encryption scheme. 

\subsection{Federated learning}\label{sec:fedavg}
Federated Learning is a distributed machine learning concept. It enables training of a machine learning model from data that is kept at decentralized devices.
As shown in Fig~\ref{fig:Comparisona}, a remotely participating device first downloads the global model from the server, trains it with local data, and then summarizes the results into an model update that is returned to the server. 
At the server, updates from remote devices are aggregated into a new global model, to be shared again.  
The server and remote devices periodically exchange model updates such as weights or gradients computed during the training.

Stochastic Gradient Descent (SGD) can be naively applied in distributed learning, but requires many communication rounds.
To address this, McMahan et al.~\cite{mcmahan2017communication} proposed FederatedAveraging (FedAvg), which exploits iterative model averaging in federated learning to reduce the communication load as follows. 
For $N$ remote devices $d_i,i\in\{1,\dots,N\}$, a fraction of these devices $\hat{N}$ performs local model training in each round. 
In particular, in each round, $\hat{N}$ remote devices $d_i$ conduct $L$ training epochs on their local datasets before sharing the model parameters ${w_{t + 1}^i}$ with the server. 
The server computes an average of all devices' model parameters as the updated global model.
This process is repeated until convergence.

\begin{algorithm}
\renewcommand{\thealgorithm}{}
\caption{FederatedAveraging(FedAvg)}\label{FedAvg}
\begin{algorithmic}[1]

\STATEx {Let $\eta $ be the learning rate, and $\mathcal{B}$ be the minibatch size for local model training.}

\STATEx {The server initializes initialize $w_0 $ // Global model parameters}
\FOR{each round $t = 1,2,...$ }
    \STATE   ${S_t} \leftarrow $ (random set of remote devices $d_i$);\$
    \FOR{each remote device $d_i \in {S_t}$ in parallel} 
        \STATE $w_{t + 1}^i \leftarrow \mbox{DeviceUpdate}(d_i,{w^i_t})$, 
        \STATE ${w_{t + 1}} \leftarrow \frac{1}{\hat{N}}\sum\limits_{i = 1}^{\hat{N}} {w_{t + 1}^i} $ 
    \ENDFOR
\ENDFOR

Remote devices compute their local updates 

\STATE $\mbox{DeviceUpdate}(d_i,{w^i_t})$: $\mathcal{B} \leftarrow $ (select batches of size $\mathcal{B}$)  //Run by each remote device $d_i$ locally
\FOR{local epochs $l=1,\dots,L$ }
    \FOR{batch $b \in \mathcal{B}$}
        \STATE $w^i_{t+1} \leftarrow w^i_t - \eta \nabla \ell (w^i_t;b)$; // update the weight using the loss $\ell (w^i_t;b)$ and the learning rate $\eta$
    \ENDFOR
\ENDFOR
\STATE return $w^i_{t+1}$ to server
 
\end{algorithmic}
\end{algorithm}

\subsection{Multi-key Homomorphic encryption}
An encryption scheme $E(k, x)$ for a key $k$ and an input $x$ is called homomorphic if for the encryption algorithm $E$ and operation $f$, there is an efficient algorithm $G$ such that
\begin{equation} 
E\left(k,f\left(x_1,\dots,x_n\right) \right) = G\left(k,f\left( E(x_1),\dots,E(x_n) \right) \right).
\label{eq_homEnc}
\end{equation}
If equation~(\ref{eq_homEnc}) only holds for either addition or multiplication, the scheme is called partially homomorphic encryption. 
It is fully homomorphic encryption, if it holds for both addition and multiplication (FHE).

Multi-key homomorphic encryption (MK-HE) allows different participants to use different keys for encryption. The aggregated ciphertext obtained after performing polynomial operations on different individual ciphertexts can only be jointly decrypted by combining the respective secret keys associated with these ciphertexts.  

MK-CKKS is a MK-HE scheme~\cite{chen2019efficient} which is a multi-key variant of the CKKS FHE scheme~\cite{cheon2017homomorphic} that supports approximate fixed-point arithmetic. 
Since homomorphic multiplication is not involved in our federated learning mechanism, we introduce the additive homomorphism of MK-CKKS only. 

MK-CKKS is a Ring Learning with Errors~\cite{lyubashevsky2010ideal} (RLWE)-based homomorphic encryption scheme. 
Let 
\begin{equation}
    R = \mathbb{Z}[X]/({X^n}+1)\nonumber
\end{equation} 
be the cyclotomic ring in which $n$ has the power of two dimension, 
and in which $\mathbb{Z}[X]$ is the polynomial ring with integer coefficients and the elements in $R$ satisfy $X^n=-1$.
${R_q} = {\mathbb{Z}_q}[X]/({X^n} + 1)$ is the residue ring of R with coefficients modulo an integer q. 
For parameters $(n,q,\chi ,\psi )$, our RLWE assumption is that, given polynomials 
of the form $(a,b = s \cdot a + e) \in R_q^2$, the term $b$ is computationally indistinguishable from uniformly random elements of $R_q$ when $a$ is chosen uniformly at random from $R_q$, $s$ is chosen from the key distribution $\chi $ over $R_q$, and $e$ is drawn from the error distribution $\psi $ over $R$~\cite{lyubashevsky2010ideal}.

We denote vectors in bold and use $\left\langle\bm{ {{u,v}}} \right\rangle $ to denote the dot product of two vectors $\bm{{{u}}}$ and $\bm{{v}}$.
$x \leftarrow \Gamma$ denotes the sampling of $x$ according to the distribution $\Gamma$. $\lambda $ denotes the security parameter throughout the paper: all known valid attacks against the cryptographic scheme under scope should take $\Omega ({2^\lambda })$ bit operations.  $g \in {Z^d}$ is an integral vector, called the gadget vector. MK-CKKS assumes the Common Reference String (CRS) model so all devices share a random polynomial vector $\bm{{a}} \leftarrow U(R_q^d)$, here $U( \cdot )$ represents the uniform distribution.
Let $\bm{{{sk}_i}} = (1,{s_i})$ for the secret key ${s_i} $, $\bm{\overline {sk} } = (1,{s_1},...{s_N})$ for the concatenation of  multiple secret keys.
Let $\bm{{c{t_i}}} = (c_0^{{d_i}},c_1^{{d_i}})$ be the ciphertext of plaintext $m_i$ from remote device $d_i$, $i=1,\dots,N$.
 \\[.02cm]

{\bfseries Setup.} For a given security parameter $\lambda $, set the RLWE dimension $n$, ciphertext modulus $q$, key distribution
$\chi $ and error distribution $\psi $ over $R$. Generate a random vector $\bm{{a}} \leftarrow U(R_q^d)$. Return
the public parameter $(n,q,\chi ,\psi ,{\bm{{a}}})$. A remote device $d_i$ generates its secret key $s_i \leftarrow \chi $, and computes its public key as ${{\bm{{b}}}_i} =  - {s_i} \cdot {\bm{{a}}} + {\bm{{e_i}}}  \in R_q^2$, here ${\bm{{e_i}}}$ is an error vector drawn from the error distribution $\psi$ over $R$.

{\bfseries Encoding and decoding.} 
Before encryption, a complex number is first expanded into a vector, and then encoded as a polynomial of ring R based on the complex canonical embedding map. 
The decoding transfers a polynomial into a complex vector after decryption.

{\bfseries Encryption.} 
After encoding a message vector into a plaintext $m_i$, that is an element of a cyclotomic ring, $d_i$ then encrypts $m_i$ as a ciphertext $\bm{{{ct}_i}} = (c_0^{d_i},c_1^{d_i})$
where ${c_0^{d_i}} = v_i \cdot b_i + m_i + e_0^{d_i}\, (\bmod \,q)$ and ${c_1^{d_i}} = v_i \cdot a + e_1^{d_i}\, (\bmod \,q)$.
Here $a = \bm{{a}}[0]$ and $b_i = {{\bm{{b}}}_i}[0]$,  $v_i \leftarrow \chi $ and ${e_0^{d_i}},{e_1^{d_i}} \leftarrow \psi $. 
Small errors are inserted to ensure security and they can be removed by the rounding operation after carrying out homomorphic operations. 
In MK-CKKS, an additive ciphertext associated to N different parties is of the form $\bm{C_{sum}} \overset{\mbox{\tiny def}}{=} \left(\sum\limits_{i = 1}^N c_0^{d_i},c_1^{d_1},c_1^{d_2},\dots,c_1^{d_N}\right) \in {R_q^{N + 1}}$.

{\bfseries Decryption of individual ciphertext.} $d_i$ computes a dot product of 
$\bm{{{sk}_i}} = (1,{s_i})$ and $\bm{{{ct}_i}} = ({c_0^{d_i}},{c_1^{d_i}})$ as follows. 
\begin{eqnarray}
{\rm{ < }}\bm{{{ct}_i}},\bm{{{sk}_i}} > \, (\bmod \,q) &=& {c_0^{d_i}} + {c_1^{d_i}} \cdot s_i \, (\bmod \,q)\nonumber\\
&=& v_i \cdot b_i + m_i + {e_0^{d_i}} + v_i \cdot a \cdot s_i + {e_1^{d_i}}\cdot s_i\, (\bmod \,q)\nonumber\\
&=& v_i \cdot ( - s_i \cdot a + e_i) + m_i + {e_0^{d_i}} + v_i \cdot a \cdot s_i + {e_1^{d_i}}\cdot s_i\, (\bmod \,q)\nonumber\\
&=& m_i + v_i \cdot e_i + {e_0^{d_i}} + {e_1^{d_i}}\cdot s_i\, (\bmod \,q)\nonumber\\
&\approx& m_i\nonumber
\end{eqnarray}

{\bfseries Additive homomorphism.} 
Let $\bm{{{ct}_i}} = (c_0^{{d_i}},c_1^{{d_i}})$ and $\bm{{{ct}_j}} = (c_0^{{d_j}},c_1^{{d_j}})$ be two ciphertexts of plaintext messages $m_i$ and $m_j$ from remote devices $d_i$ and $d_j$. 
The sum of the ciphertexts is $\bm{C_{sum}} \overset{\mbox{\tiny def}}{=} (c_0^{{d_i}} + c_0^{{d_j}},c_1^{{d_i}},c_1^{{d_j}})$. 
It can be decrypted by computing a dot product of $\bm{C_{sum}}$ and $\bm{\overline {sk} } = (1,s_i,s_j)$. 
The correctness is proved as follows:
\begin{eqnarray}\nonumber
{\rm{ < }} \bm{C_{sum}} ,\bm{\overline {sk} } > \, (\bmod \,q) &=& (c_0^{{d_i}} + c_0^{{d_j}}) + c_1^{{d_i}} \cdot {s_i} + c_1^{{d_j}} \cdot {s_j}\, (\bmod \,q)\\
\nonumber &=& (c_0^{{d_i}} + c_1^{{d_i}} \cdot {s_i}) + (c_0^{{d_j}} + c_1^{{d_j}} \cdot {s_j})\, (\bmod \,q)\\
\nonumber&\approx& {m_i} + {m_j}
\end{eqnarray}

{\bfseries Decryption of sum.} The distributed decryption based on noise flooding is introduced in MK-CKKS since it is not reasonable to assume that any party holds multiple secret keys. 
The decryption consists of two algorithms: partial decryption and merge.
\begin{description}
\item[$\bm{{\rm{MK-CKKS}}{\rm{.PartDec}}}(c_1^{{d_i}}{\rm{,}}{s_i}{\rm{)}}$:] Given a polynomial $c_1^{{d_i}}$ and a secret ${s_i}$, sample
an error $e_i^* \leftarrow \phi $ and return ${\mu _i} = c_1^{{d_i}} \cdot {s_i} + e_i^* \, (\bmod \,q) $.

\item[$\bm{{\rm{MK-CKKS}}{\rm{.Merge}}}(\sum\limits_{i = 1}^N {c_0^{{d_i}}} {\rm{,}}{\{ {\mu_i}\}_{1 \le i \le N}}{\rm{)}}$:] Compute and return $\mu {\rm{ = }}\sum\limits_{i = 1}^N {c_0^{{d_i}}}  + \sum\limits_{i = 1}^N {{\mu _i}}  \, (\bmod \,q) \approx {\rm{ < }} \bm{C_{sum}} ,\bm{\overline {sk} } > \, (\bmod \,q) $.
\\[.08cm]
Here $e_i^*$ is generated from error distribution $\phi $ which has a larger variance than the standard error distribution $\psi$.
\end{description}

\section{Multi-key homomorphic encryption for Federated learning}
MK-CKKS is not directly applicable to federated learning, since the server would be capable of decrypting the individual model updates and hence learn about the private data. 
We improve the MK-CKKS protocol in this respect so that the server becomes only able to decrypt the aggregation of the shared encrypted model updates, thereby obfuscating the individual model updates with updates from all other remote participating devices.  
The proposed xMK-CKKS simplifies the decryption process of MK-CKKS while maintaining strong privacy protection. 
Based on xMK-CKKS, we propose a privacy-preserving federated learning scheme.

\subsection{Threat model}
Most current cryptographic protocols assume honest-but-curious participants, which means that the participants execute the protocol honestly, but exploit any opportunity to extract private data from intermediate results generated during the execution of the cryptographic protocol. 

In our work, we apply homomorphic encryption in a federated learning scenario to protect data privacy. 
In this context, we assume that the server and all remotely participating devices are honest-but-curious. 
This means that they follow the scheme honestly, but have an intention of inferring the private information of other devices from the information shared during the execution of the protocol.
We further assume that collusion may exist between devices and the server. 
In particular, we consider an attack in which $k<N-1$ devices collude together with the server, to jointly attack particular devices, where $N$ is the total number of devices and k is the number of colluding devices.

\subsection{xMK-CKKS}
As described in Section~\ref{sec:preliminaries}, the decryption of the sum of the ciphertexts requires two steps, partial decryption and merge. 
In a federated learning scenario, each device sends encrypted model updates to the server for aggregation. 
In this way, the server obtains $\bm{{ct_i}} = (c_0^{d_i},c_1^{d_i})$. 
If $\mu_i = c_1^{d_i} \cdot s_i + e_i^*\, (\bmod \,q)$ were also to be shared, the server could directly decrypt $m_i$ (cf. equation~(\ref{eq:directDecrypt})).
\begin{equation}
c_0^{d_i} + \mu _i = c_0^{d_i} + c_1^{d_i} \cdot s_i + e_i^* \approx m_i\, (\bmod \,q)\label{eq:directDecrypt}
\end{equation}

To resolve this problem, the server shall not obtain $c_0^{d_i}$ and $\mu _i$ at the same time. 
For instance, either $\sum\limits_{i = 1}^N c_0^{d_i} $ or $\sum\limits_{i = 1}^N \mu _i$ may instead be computed collaboratively by the remote devices. 
However, this would introduce the risk that private information $m_i$ leaks if devices and the server collude. 

{\bfseries xMK-CKKS}. 
To avoid privacy leakage during decryption, we propose xMK-CKKS. 
In xMK-CKKS, an aggregated public key is computed for encryption by aggregating the public keys of all devices. 
For decryption, the server requires a decryption share $D_i$ computed by each device $d_i$. 
$D_i$ combines the sum over all cipertexts as well as the secret key $s_i$ and error term 
$e_i^*$ of $d_i$ (see equation~(\ref{eq:decryptionShare})). 
The following methods define the xMK-CKKS scheme in detail.

\begin{description}
\item[${\rm{Setup}}(1^\lambda)$:] Given a security parameter $\lambda $, set the RLWE dimension $n$, ciphertext modulus $q$, key distribution $\chi$ and error distribution $\psi $ and $\phi$ over $R$. 
Generate a random vector  $\bm{{a}} \leftarrow U(R_q^d)$. Return the public parameters $(n,q,\chi ,\psi ,\phi, \bm{{a}})$.

\item[${\rm{KeyGen}}(n,q,\chi ,\psi ,\bm{{a}})$:] Each device $d_i$ ($1 \le i \le N$) generates its secret key $s_i$ and computes its public key  
${{\bm{{b}}}_i} =  - {s_i} \cdot {\bm{{a}}} + {\bm{{e_i}}} \,(\bmod \,q)$. 
We define the aggregated public key $\bm{\widetilde {{b}}}$ as 
\begin{equation}
\bm{\widetilde {{b}}} = \sum\limits_{i = 1}^N {\bm{{b}}}_i  = \sum\limits_{i = 1}^N ( - {s_i}) \cdot\bm{{a}} + \sum\limits_{i = 1}^N {\bm{{e_i}}} \, (\bmod \,q)
\end{equation}

\item[${\rm{Enc}}(m_i,\bm{\widetilde {{b}}},\bm{{a}})$:] The plaintext $m_i$ of a device $d_i$ is encrypted as
\begin{equation}
\bm{{{ct}_i}} = (c_0^{d_i},c_1^{d_i}) = (v^{d_i}\cdot \widetilde {{b}} + m_i + e_0^{d_i},v^{d_i}\cdot {a} + e_1^{p_i}){\mkern 1mu}\,(\bmod \,q)
\end{equation}

Here ${a} = \bm{{a}}[0]$, $\widetilde {{b}} = \bm{\widetilde {{b}}}[0]$, $v \leftarrow \chi $, $e_0^{d_i},e_1^{d_i} \leftarrow \psi $.

\item[${\rm Add}(\bm{{{ct}_1}},\dots,\bm{{{ct}_N}})$:] The sum of all ciphertexts is
\begin{eqnarray}
\bm{C_{sum}} &=& \sum\limits_{i = 1}^N \bm{{{ct}_i}} \, \mathop  = \limits^\Delta \, (C_{sum_0},C_{sum_1}) \nonumber\\
  &=& \left( \sum\limits_{i = 1}^N c_0^{d_i} ,\sum\limits_{i = 1}^N c_1^{d_i} \right) \nonumber\\
  &=& \left( \sum\limits_{i = 1}^N (v^{d_i}\cdot \widetilde {{b}} + m_i + e_0^{d_i}) ,\sum\limits_{i = 1}^N (v^{d_i}\cdot {a} + e_1^{d_i}) \right)\,(\bmod \,q)
\end{eqnarray}

\item[${\rm Dec}(\bm{C_{sum}},s_1,\dots,s_N)$:] Each device $d_i$ computes its decryption share $D_i$
\begin{eqnarray}
D_i &=& s_i\cdot C_{sum_1} + e_i^* \nonumber\\
    &=& s_i\cdot\sum\limits_{i = 1}^N (v_i\cdot {a} + e_1^{d_i} ) + e_i^*\,(\bmod \,q)\label{eq:decryptionShare}
\end{eqnarray}

Here $e_i^*\leftarrow \phi$.

Then, the sum of all plaintexts can be recovered as follows.
\noindent\begin{eqnarray}
C_{sum_0} + \sum\limits_{i = 1}^N D_i\,\bmod \,q\nonumber
 &=& C_{sum_0} + \sum\limits_{i = 1}^N s_i  \cdot C_{sum_1} + \sum\limits_{i = 1}^N e_i^*\,\bmod \,q\nonumber\\
 &=& \sum\limits_{i = 1}^N (v^{d_i} \cdot \widetilde {{b}} + m_i + e_0^{d_i})  + \sum\limits_{i = 1}^N s_i  \cdot \sum\limits_{i = 1}^N (v^{d_i} {a} + e_1^{d_i})  + \sum\limits_{i = 1}^N e_i^* \,\bmod \,q\nonumber\\
 &=&  - \sum\limits_{i = 1}^N v^{d_i} s_i {a} + \sum\limits_{i = 1}^N v^{d_i}\sum\limits_{i = 1}^N e_i + \sum\limits_{i = 1}^N m_i + \sum\limits_{i = 1}^N e_0^{d_i}  + \sum\limits_{i = 1}^N v^{d_i} s_i  {a} + \sum\limits_{i = 1}^N \left(s_i e_1^{d_i}  + e_i^*\right) \,\bmod \,q\nonumber\\
 &=& \sum\limits_{i = 1}^N m_i  + \sum\limits_{i = 1}^N v^{d_i}  \cdot \sum\limits_{i = 1}^N e_i  + \sum\limits_{i = 1}^N e_0^{d_i}  + \sum\limits_{i = 1}^N \left(s_i e_1^{d_i} + e_i^*\right) \,\bmod \,q\nonumber\\
 &\approx& \sum\limits_{i = 1}^N m_i
\end{eqnarray}
\end{description}

In xMK-CKKS, the aggregated public key is computed for encryption. 
The decryption requires each device to computes its decryption share. The decryption share implicitly contains the information of individual secret key of each participants and the aggregated cipheretexts $\bm {C_{sum}}$, and adds an error for security, thus is useless to decrypt any other ciphertext including individual ciphertext.
Therefore, the risk of privacy leakage during decryption is mitigated by the protocol. 
Consequently, in federated learning and similar collaborative learning scenarios scenarios, xMK-CKKS provides stronger security than MK-CKKS. 
We show in section~\ref{sec:security}, that it is also robust against any collusion between $k < N-1$ honest-but-curious devices and the server. 
Note that xMK-CKKS doesn’t require any interaction 
among devices, and is therefore suited also to scenarios in which devices are not fully interconnected. 

\subsection{Privacy-preserving federated learning based on xMK-CKKS}

We propose a privacy-preserving federated learning scheme based on xMK-CKKS.
As federated learning mechanism, we apply the FedAvg scheme (cf. section~\ref{sec:fedavg}) with the fraction of remotely participating devices as $1$ (all devices contribute to model training in each round). 

\begin{figure}
	\centering
	\includegraphics[scale=0.6]{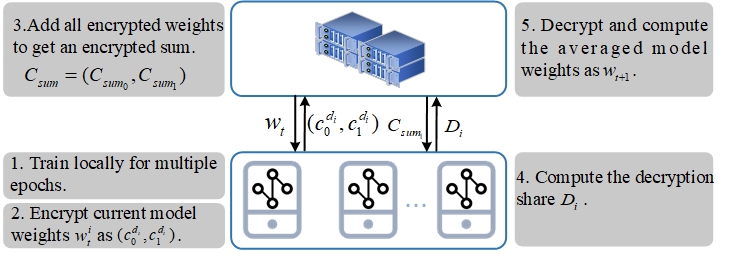}
	\caption{Privacy-preserving federated learning based on xMK-CKKS multi-key homomorphic encryption}
	\label{fig:Overview}
\end{figure}

The complete model training process consists of multiple aggregation rounds between the server and the devices. Fig~\ref{fig:Overview} details the process of one aggregation round. In each aggregation round $t$, all devices obtain the current global model from the server, train it for multiple epochs on local data (step 1 in figure~\ref{fig:Overview}) and encrypt the resulting local model weights with the aggregated public key $\bm{\widetilde {{b}}}$ (step 2). The ciphertexts are then sent by all remote devices to the server where they are aggregated as the sum over all encrypted model weights (step 3). In order to obtain the plaintext from this encrypted sum, the server requires all devices to compute their decryption shares $D_i$ (step 4). After sharing the decryption shares, the server merges them with the ciphertext to decrypt the encrypted sum over all shared weighted model weights (step 5). In particular, after decryption, the server obtains the sum of all devices' weights $\sum\limits_{i = 1}^N w_t^i$ before computing averaged weights ${w}_{{t+1}}$ as the new model weights for next aggregation round. This procedure is iterated until model convergence. We detail the individual steps in the following. The steps 1 to 5 are executed iteratively until model convergence.\\[.02cm]

\noindent{\bfseries Setup:}
For a given {security parameter} $\lambda$, set the RLWE dimension $n$, ciphertext modulus $q$, key distribution $\chi$ and error distribution $\psi $ and $\phi$ over $R$. 
Generate a random vector $\bm{{a}} \leftarrow U(R_q^d)$. Return the public parameters $(n, q, \chi, \psi,\phi, \bm{{a}})$. 
Each device $d_i$ samples a secret key $s_i \leftarrow \chi $ and computes its public key ${{\bm{{b}}}_i} =  - {s_i} \cdot {\bm{{a}}} + {\bm{{e_i}}}\, (\bmod \,q)$.
Then all participating devices collaboratively compute the aggregated public key as 
\begin{equation}
\bm{\widetilde {{b}}} = \sum\limits_{i = 1}^N {\bm{{b}}}_i  = \sum\limits_{i = 1}^N ( - {s_i}) \cdot\bm{{a}} + \sum\limits_{i = 1}^N {\bm{{e_i}}} \, (\bmod \,q)
\end{equation}

\noindent{\bfseries Step 1: Local training}:
In each aggregation round $t$, each device $d_i$ first downloads the current global model weights $w_t$ from the server. 
Every device applies model optimization, for instance, SGD or Adam, to train the obtained global model on the local data. 
After multiple training epochs, each device derives a local model with weights $w_t^i$.

\noindent{\bfseries Step 2: Encryption of model weights}:
Let $m_i \in R$ be an encoded plaintext input of $w_t^i$ and let ${a} = \bm{{a}}[0]$, $\widetilde {{b}} = \bm{\widetilde {{b}}}[0]$. 
Sample $v^{d_i} \leftarrow \chi $ and $e_0^{d_i},e_1^{d_i} \leftarrow \psi $. 
Compute the ciphertext \begin{equation}
\bm{{ct^{d_i}}} = (c_0^{d_i},c_1^{d_i}) = (v^{d_i}\cdot \widetilde {{b}} + m_i + e_0^{d_i},v^{d_i}\cdot {a} + e_1^{d_i}){\mkern 1mu}\,(\bmod \,q)
\end{equation}
Then, each device $d_{i}$ shares $\bm{{ct^{d_i}}}=(c_0^{d_i},c_1^{d_i})$ with the server.

\noindent{\bfseries Step 3: Computation of the homomorphic sum}:
After receiving encrypted weights from all devices, the server aggregates them by adding them together to get an encrypted sum $\bm{C_{sum}}$. 
The server then publishes $C_{sum_1}$ for all devices.

\begin{equation}
\bm{C_{sum}} = \sum\limits_{i = 1}^N \bm{{ct^{d_i}}}
= \left(\sum\limits_{i = 1}^N c_0^{d_i} ,\sum\limits_{i = 1}^N c_1^{d_i} \right)\mathop  = \limits^\Delta  (C_{sum_0},C_{sum_1})\,(\bmod \,q)
\end{equation}

\noindent{\bfseries Step 4: Computation of the decryption share}:
All the devices $d_i$ 
are required to decrypt the multi-key ciphertext. 
For this, each device computes its decryption share $D_i$ and sends it to the sever. 
We set the noise $e_i^*$ as $e_i^* \leftarrow \phi $ and assume that the distribution $\phi$ has a larger variance than the standard error distribution $\psi$ of the basic scheme.
\begin{equation}
D_i = s_i\cdot C_{sum_1} + e_i^* = s_i\cdot \sum\limits_{i = 1}^N (v_i\cdot {a} + e_1^{d_i} ) + e_i^*\,(\bmod \,q)
\end{equation}

\noindent{\bfseries Step 5:  Model aggregation}:
After receiving all decryption shares, the server merges all decryption shares $D_i, i\in\{1,\dots,N\}$ with $C_{sum_0}=\sum\limits_{i = 1}^N c_0^{d_i} $ to recover the plaintext.
\begin{equation}
\sum\limits_{i = 1}^N m_i  \approx C_{sum_0} + \sum\limits_{i = 1}^N D_i \,(\bmod \,q)
\end{equation}

Finally, the server decodes $\sum\limits_{i = 1}^N m_i $ to obtain the sum over the weights $\sum\limits_{i = 1}^N w_t^i$ before computing the averaged weights $w_{t + 1}$ as the updated model weights for next round. 
\begin{equation}
w_{t + 1} = \frac{1}{N}\sum\limits_{i = 1}^N w_t^i \mkern 1mu 
\end{equation}

\section{Security analysis}
\label{sec:security}
We discuss how our scheme guarantees privacy for the data hosted at distributed devices by ensuring the confidentiality of the model weights. 
In particular, the following theorems describe the security of the scheme with respect to various potential adversaries. 

{\bfseries Theorem 1. Security against an honest-but-curious Server}: In our scheme, an honest-but-curious server can not infer any private information about the devices' data.

\textit{Proof}: In our xMK-CKKS based federated learning scheme, remotely participating devices $d_i$ send two types of information to the server. 
First, in step 2, $d_{i}$ shares ciphertext $\bm {{ct^{d_i}}}$ with the server, which is encrypted by xMK-CKKS. 
Then, in step 4, $d_{i}$ sends its decryption share $D_i$ to the sever. 
Here we have $\bm{{ct^{d_i}}} = (c_0^{d_i},c_1^{d_i})$,
where $c_0^{d_i} = v^{d_i}\cdot \widetilde b + m_i + e_0^{d_i}\,(\bmod \,q)$ and $c_1^{d_i}= v^{d_i}\cdot {a} + e_1^{d_i}\,(\bmod \,q)$, $D_i = s_i \cdot C_{sum_1} + e_i^* = s_i\cdot\sum\limits_{i = 1}^N (v_i\cdot {a} + e_1^{d_i} ) + e_i^*\,(\bmod \,q)$. 
All messages contain an added error to guarantee the security according to RLWE assumption. 
RLWE guarantees that the $c_0^{d_i}$ and $D_i$ are computationally indistinguishable from uniformly random elements of $R_q$. 
They don't leak any information of $m_{i}$ and $s_{i}$ to the server $S$. 
After collaborative decryption, the server can only get a sum of all gradients, which also does not leak information on the individual weights.  

Therefore, our scheme can ensure the confidentiality of individual weights, and thereby guarantee the privacy of the data distributed at remote devices. 
The server cannot infer any private information about the device from the information it receives.

{\bfseries Theorem 2. Security against honest-but-curious distributed devices}: 
In our scheme, an honest-but-curious device can not infer any private information by stealing the shared information of other devices.

\textit{Proof}: In our scheme, the model updates of each device $d_i$ are encrypted via xMK-CKKS which is based on RLWE. 
Each device individually chooses a secret key $s_i$ to establish a public key $\bm {{b_i}}$. 
In addition, all devices collaboratively compute an aggregated public key to encrypt their weights. 
An error is added to the decryption share to protect the secret key of each device. 
Therefore, an honest-and-curious device can not infer any information by stealing the information uploaded by other devices. 

{\bfseries Theorem 3. Security against the collusion between devices and the server}: Collusion between $k<N-1$ devices and the server does not leak information about model updates from other devices, where $N$ is the total number of devices and k is the number of colluding devices

\textit{Proof}: In our scheme, model updates of each device are encrypted by the aggregated public key $\bm{\widetilde {{b}}} = \sum\limits_{i = 1}^N {\bm{{b}}}_i  = \sum\limits_{i = 1}^N ( - {s_i}) \cdot\bm{{a}} + \sum\limits_{i = 1}^N {\bm {{e_i}}} \, (\bmod \,q)$ and shared with the server. 
Then, the server computes a sum over the model weights. 
The decryption of the individual encrypted weights $\bm{{ct^{d_i}}}$ and the encrypted sum $\bm{C_{sum}}$ can only be done by combining the decryption key shares of all devices. 

The first collusion attack is that colluding parties try to infer $m_i$ of the victim $d_i$ from its individual ciphertext $\bm{{ct^{d_i}}}$. 
In the worst case, we assume a victim $d_i$ while the other $N-1$ devices collude with the server $S$. 
A possible attempt is to compute $c_1^{{d_j}} \cdot {s_j}$ for $j \not= i$, then to merge them with $c_0^{{d_i}}$. 
\noindent\begin{eqnarray}
c_0^{{d_i}} + \sum\limits_{j \not= i}^N {c_1^{{d_1}} \cdot {s_i}} \, (\bmod \,q)\nonumber 
&=& {v^{{d_i}}}\cdot \mathop b\limits^ \sim   + {m_i} + e_0^{{d_i}} + \sum\limits_{j \not= i} {({v^{{d_j}}}\cdot a + e_1^{{d_j}}) \cdot {s_j}} \, (\bmod \,q)\nonumber\\
&=&  - \sum\limits_{i=1}^N {{v^{{d_1}}} \cdot {s_i} \cdot a}  + {m_i} + e_0^{{d_i}} + \sum\limits_{j \not= i} {{v^{{d_j}}}\cdot{s_j} \cdot a + \sum\limits_{j \not= i} {e_1^{{d_j}} \cdot {s_j}} } \, (\bmod \,q)\nonumber\\
&=& - {v^{{d_i}}}\cdot{s_i} \cdot a + {m_i} + e_0^{{d_i}} + \sum\limits_{j \not= i} {e_1^{{d_j}} \cdot {s_j}}
\end{eqnarray}
The result is roughly equivalent to a partial ciphertext encrypted by the individual public key $b_i$, therefore has no privacy risk. Even if the other devices collude and are in possession of the secret keys $s_j,j\not=i$, they are not able to jointly decrypt the individual ciphertext encrypted with $b_i$ by the victim $d_i$. 

We note thought that the colluding parties might try to infer $m_i$ from the decrypted aggregation result $\sum\limits_{i = 1}^N {{m_i}} $. 
We argue that the attack cannot be successful as long as there are two uncompromised parties. 
In the worst case, the $N-2$ colluding parties subtract their plaintext $m_i$ from the sum $\sum\limits_{i = 1}^N {{m_i}} $. 
The result is the sum of plaintexts of two uncompromising parties, therefore have no privacy risk for individual data.

Therefore, our protocol can resist collusion between $k<N-1$ devices and the server.

\section{Evaluation}
In this section, we evaluate the propose xMK-CKKS based federated learning scheme on a smart healthcare scenario, in particular, in elderly-fall detection. 
Specifically, we compare three schemes to it: traditional federated learning, federated learning based on MK-CKKS (https://github.com/snucrypto/HEAANe) and federated learning based on Paillier homomorphic encryption (http://hms.isi.jhu.edu/acsc/libpaillier/). 
We then report fall detection model accuracy, energy consumption, communication cost, as well as computational cost during the whole training process.

\subsection{Dataset}
We focus on fall detection for elderly persons as an universally relevant problem related to health and general wellbeing. 
In particular, we employ a multimodal dataset called "UP-FALL"~\cite{UP-FALL_dataset} which includes 17 healthy young individuals. Each subject performs 10 different activities and falls with three trials each.
The data is collected from wearable and ambient sensors, and vision devices.
The subjects wear sensors on the body to collect sensor data. 
The model is trained to recognize five types of falls,  1) \textit{falling forward using hands}, 2) \textit{falling forward using knees}, 3) \textit{falling backward}, 4) \textit{falling sitting in an empty chair}, and 5) \textit{falling sideward}. 
If a person falls, the model shall detect this.

In our evaluation, we utilize the dataset from the data of the accelerometer sensors of 10 devices. 
We allocate one Jetson Nano IoT device with the data of one acceleration sensor. 
In table~\ref{table:data_dist}, we detail the size of data distributed to each device which is on average approximately 25,000 samples per device. 
To evaluate the accuracy of the model, we construct a global testing set consisting of subset of around 37,000 samples where
no redundancy to data used for the global testing set and with the data of the distributed devices.

\begin{table}
 \caption{Data Distribution for each distributed device in acceleration samples. The testset is jointly utilized by all devices.}
  \centering
  \begin{tabular}{ cccccccccccc}
   \toprule
    Node&N1&N2&N3&N4&N5&N6&N7&N8&N9&N10&Testset\\
    \midrule
    Data size  &28K &27K&28K &27K &28K&25K&28K&26K&27K&27K& 37K \\
    \bottomrule
  \end{tabular}
  \label{table:data_dist}
\end{table}

\begin{figure}
    \centering
    \subfloat[Distribution strategy to remote devices. The server shares a small fraction of the data to decrease weights  divergences]{\includegraphics[width=0.49\textwidth]{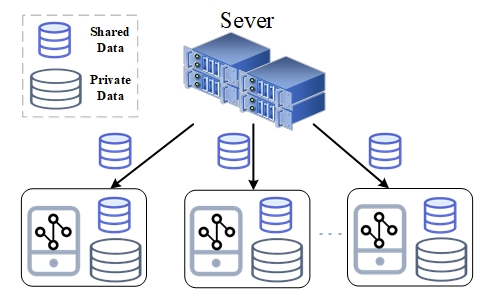}\label{fig:testbeda}} 
    \subfloat[Environment testbed settings with 10 Jetson Nano IoT devices]{\includegraphics[width=0.49\textwidth]{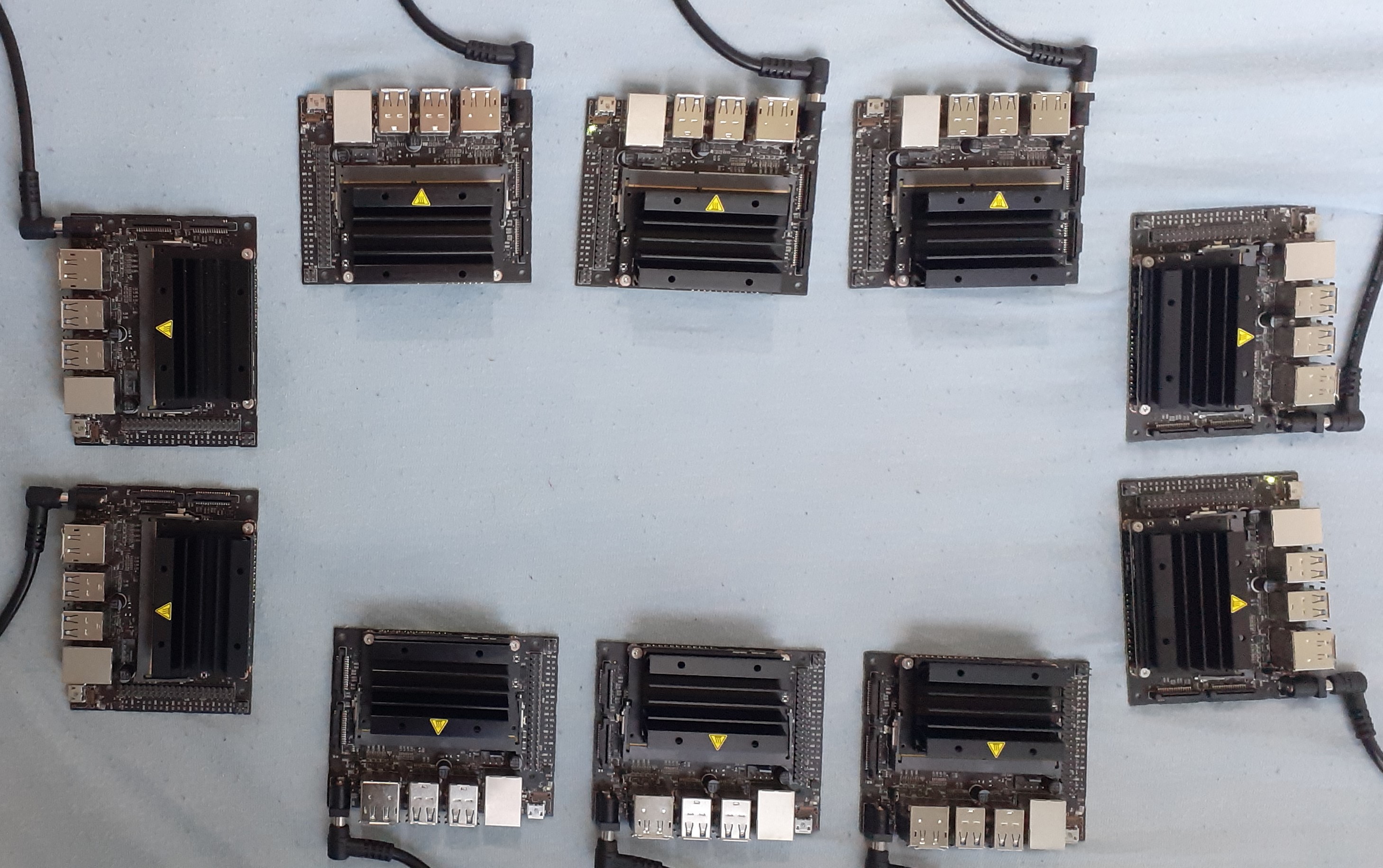}\label{fig:testbedb}} 
    
    \caption{Illustration of the experimental setting and distribution strategy utilized in our case study. 10 Jetson Nano IoT devices are utilized as remote devices storing the data while a server was computing the model aggregates.}
    \label{fig:testbed}
\end{figure}

In practice the data generated by IoT devices is exclusive, and moving data to other parties leaks private information.
The accuracy of federated learning with not independent and identically distributed (non-iid) data significantly decreases due to weight divergence. 
For our scheme, share a small amount of data, called the `global set' with all devices as suggested in~\cite{zhao2018federated} (Fig~\ref{fig:testbeda} depicts the process).

\subsection{Experimental Setup}
We set up 10 Jetson Nano nodes as our IoT devices, with 128 NVIDIA CUDA® cores, Quad-core ARM CortexA57 MPCore processor, 4 GB 64-bit LPDDR4 RAM, and 64 GB of storage (see figure~\ref{fig:testbedb}).
The Jetson Nano IoT devices are compatible with various machine learning frameworks and mainly used for model training. 
We employ as a server a laptop of Intel(R) Core(TM) i7-7700HQ CPU @ 2.80GHz (8 CPUs), ~2.8GHz. We use the TensorFlow 2.0 framework and Keras to build our CNN baseline. 
Our model consists of two-layers with ReLu activation function. 
The output layer uses softmax activation functions. 
Our model was trained with Adam optimizer~\cite{kingma2014adam} at a learning rate of $0.01$ with $20$ and $40$ local epochs in one aggregation round.

{\bfseries Comparators}. 
To validate the potency of our scheme, we implement a traditional federated learning scheme, a federated learning scheme based on MK-CKKS, and a federated learning scheme based on Paillier to compare them to our xMK-CKKS based federated learning.
The traditional federated learning has no additional privacy protection for model updates.
Federated learning based on MK-CKKS encrypts model updates by MK-CKKS, but has privacy risks due to the noise flooding technique for decryption. 
Paillier is an additively homomorphic encryption scheme which can be used in federated learning. Different from the privacy-preserving deep learning scheme proposed by Le et al.~\cite{aono2017privacy} which uses the Paillier to encrypt the gradients, we use Paillier to encrypt model weights. 
However, in Paillier, all devices share the same secret key and public key, which compromises privacy.

\subsection{Discussion of results}
We evaluated fall detection model accuracy to ensure the effectiveness of our scheme, and compared it to the Federated learning scheme without encryption. 
In particular, we analyzed the communication cost, and test the aggregation rounds with different number of local training epochs ($L$). 
In addition, since our scheme is deployed on IoT devices with constrained hardware resources, we evaluated the energy consumption of it and compared it to other federated learning schemes to validate its applicability for IoT devices. 
We further evaluated the computational cost of each scheme during encryption, decryption, computation of cipher sums as well as decryption share computation. 
The classification accuracy, communication cost, energy consumption, and computational cost are reported. 

\begin{figure}[htbp]
    \centering
    \subfloat[Accuracy of the federated lerning schemes utilizing  $L=20$ local training epochs]{\includegraphics[width=0.49\textwidth]{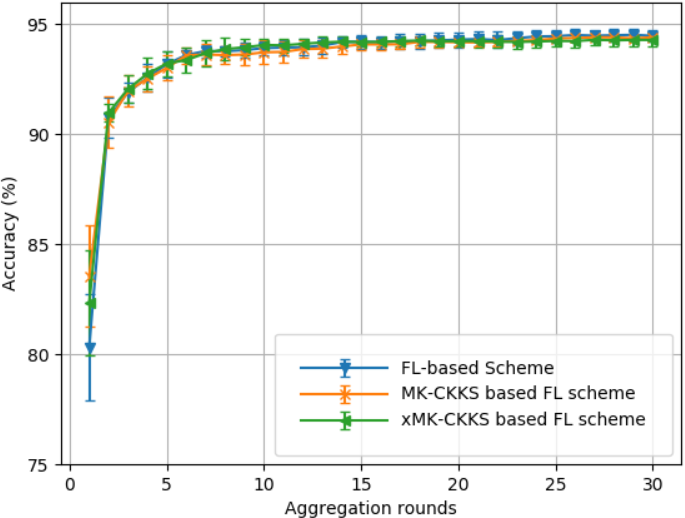}\label{fig:accuracy_alla}} 
    \subfloat[Accuracy of the federated lerning schemes utilizing  $L=40$ local training epochs]{\includegraphics[width=0.49\textwidth]{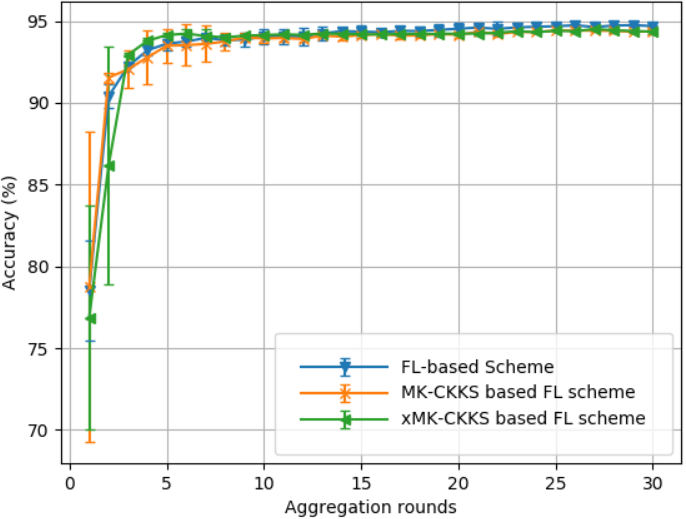}\label{fig:accuracy_allb}} 
    \caption{Classification accuracy comparison with different number of training epochs executed locally($L$) by each device in one aggregation round}
    \label{fig:accuracy_all}
\end{figure}

{\bfseries Accuracy}. We compared the accuracy of our scheme to traditional federated learning and xMK-CKKS based schemes (cf.~Fig.~\ref{fig:accuracy_all}). 
For each scheme, we executed five trials and reported the accuracy along with the standard deviation. 
When the local epochs $L=20$, the xMK-CKKS based scheme provides 93.37\% of accuracy which is almost the same as the federated learning scheme which has 93.80\% of accuracy. When we increase the local epochs to $L=40$, the xMK-CKKS based scheme provides 93.53\% of accuracy which is very close to the federated learning scheme which has 93.81\% of accuracy. The results demonstrate the accurate decryption of the xMK-CKKS scheme.

{\bfseries Communication cost}.
The fall detection model has 492 model weights and each weight is represented by 64 bits. 
In xMK-CKKS based federated learning, the ciphertext size of the model weights is approximately $87$KB. 
This is much smaller than the ciphertext size of the Paillier based Federated learning,  which is $296$KB. 
Although xMK-CKKS based federated learning needs one more communication rounds for collaborative decryption, the additional message size is only $43$KB for $C_{sum_1}$ and $43$KB for $D_i$ in each aggregation round. 
To reduce the communication cost in general, we run $20$ epochs (Fig.\ref{fig:accuracy_alla}) and $40$ epochs (Fig.\ref{fig:accuracy_allb}) of model optimization in each aggregation round. The global model converges with only $10$ or $5$ aggregation rounds. The results demonstrate that when the number of local training epoch $L$ increases, the number of aggregation rounds required for convergence decreases, thus reduce the communication cost.

{\bfseries Energy Consumption}.
We monitored the energy consumption during the operation of the IoT devices for MK-CKKS, xMK-CKKS, Paillier based federated learning schemes, federated learning without encryption, and when the IoT device is idle (Table ~\ref{table:energy}). 
The energy consumption is sampled every two seconds and we obtain the average consumption from each five of these measurements. 
The energy consumption of xMK-CKKS based federated learning is around $2.4$ watts, which is only 24\% of the maximum energy of the Jetson Nano IoT device (10 Watts). 
Therefore, we conclude that the scheme is applicable for IoT devices. 

\begin{table}
 \caption{Energy consumption of different federated learning (FL) schemes executed on distributed Jetson Nano IoT devices.}
  \centering
  \setlength\tabcolsep{3pt}
  \begin{tabular}{ cccccc}
   \toprule
    Scenario& xMK-CKKS based FL& MK-CKKS based FL &Paillier based FL&FL w/o encryption&No activity (idle)\\
    \midrule
    Energy (W) &2.4 &2.4&2.1&2.3 &1.8\\
    \bottomrule
  \end{tabular}
  \label{table:energy}
\end{table}

\begin{figure}
    \centering
    \subfloat[Encryption cost for different number of weights]{\includegraphics[width=0.24\textwidth]{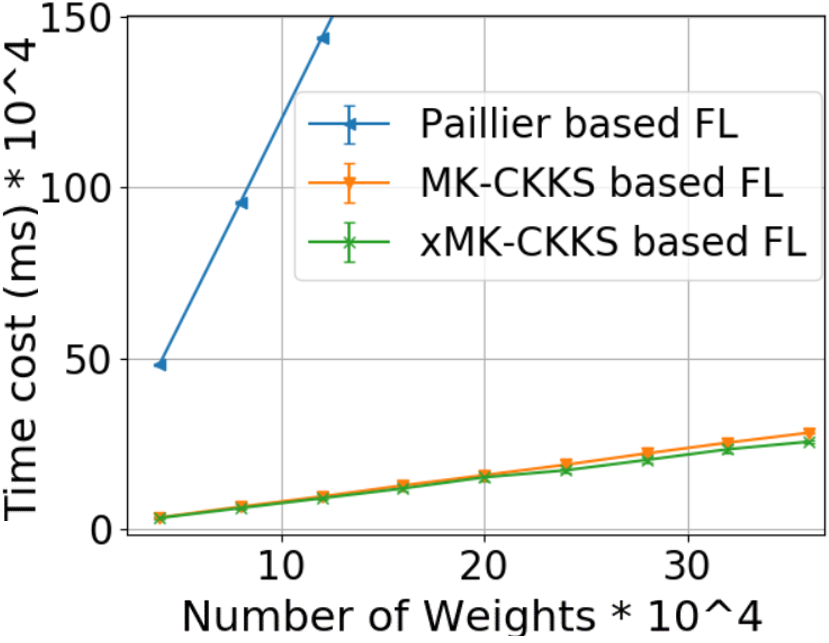}} 
    \subfloat[Decryption cost for different number of weights]{\includegraphics[width=0.24\textwidth]{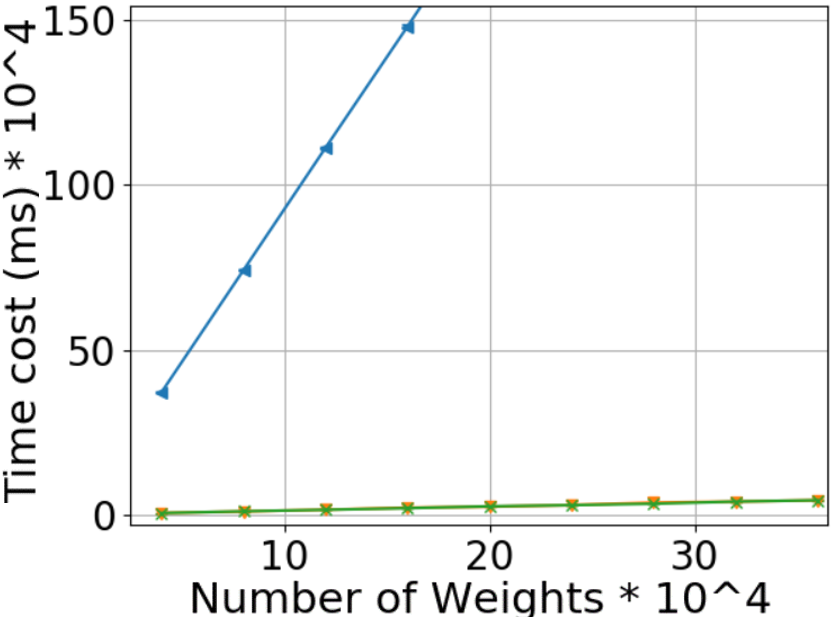}} 
    \subfloat[Ciphers sum cost for different number of weights]{\includegraphics[width=0.24\textwidth]{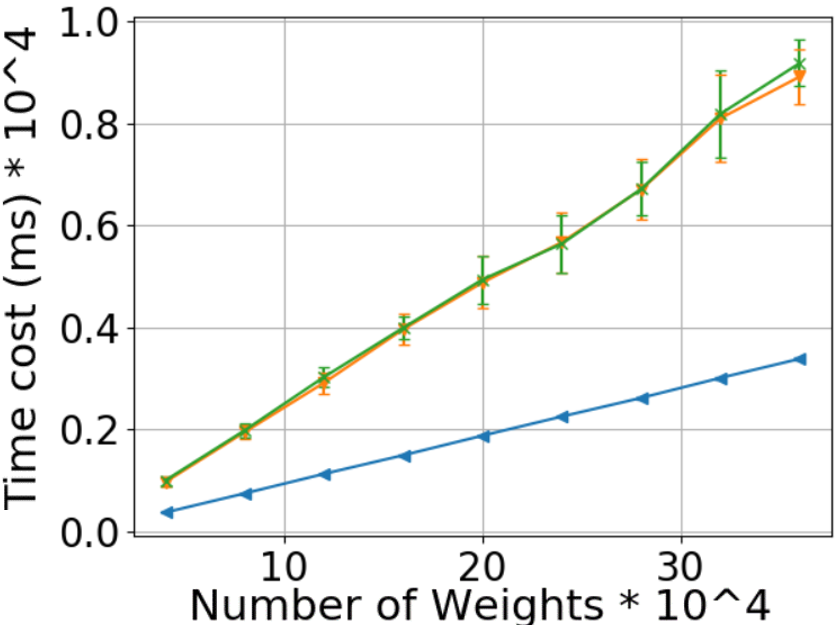}}
    \subfloat[Decryption share cost for different number of weights]{\includegraphics[width=0.23\textwidth]{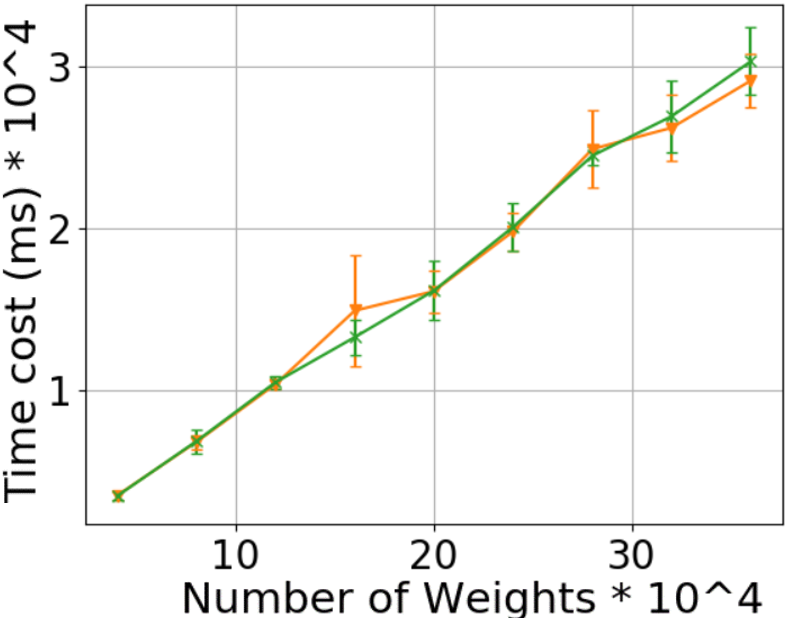}}
    \caption{Computational cost for Paillier based federated learning, MK-CKKS based federated learning, and xMK-CKKS based federated learning in different stages}
    \label{fig:results_all}
\end{figure}

{\bfseries Computational cost}
We discuss the computational cost when varying the number of weights (up to nearly 1/3 million) in Fig.~\ref{fig:results_all}. 
We report computational cost at the encryption phase~(Fig.~\ref{fig:results_all}(a)), aggregation phase~(Fig.~\ref{fig:results_all}(b)), decryption share calculation phase~(Fig.~\ref{fig:results_all}(c)), and decryption phase~(Fig.~\ref{fig:results_all}(d)),
where each experiment is executed four times.
We compare our scheme with MK-CKKS based scheme and Paillier based scheme. We notice that our scheme behaves approximately similar to MK-CKKS based scheme while significantly taking less time than the Paillier based scheme.
The results shows that our proposed scheme can be deployed in large scale data scenarios due to its reduced time cost during all phases of the scheme.

\section{Conclusion and future work}
We have proposed a novel privacy-preserving federated learning scheme based on multi-key homomorphic encryption to protect data privacy. 
In particular, we improved the MK-CKKS homomorphic encryption scheme, which has the risk of privacy leakage when it is used in the federated learning scenario to protect privacy of individual model updates.
To this end, we introduced xMK-CKKS by defining the aggregated public key and the decryption share to achieve secure and simple encryption and decryption, which is more suitable for privacy protection in federated learning scenarios. The proposed xMK-CKKS based federated learning scheme guarantees confidentiality of model updates by applying multi-key homomorphic encryption, and is robust to collusion between $k<N-1$ participating devices and the server. Therefore, the proposed scheme provides strong privacy protection for federated learning.

We evaluated this xMK-CKKS based federated learning scheme in terms of accuracy, communication cost, energy consumption, and computational cost and compared it with federated learning based on MK-CKKS homomorphic encryption, and Paillier homomorphic encryption. 
In particular, in an experiment with 10 Jetson Nano IoT devices and a single dedicated server, we executed a large-scale data scenario situated in the domain of elderly care. 
We extensively evaluated and compared the schemes. 
The experiment demonstrates that xMK-CKKS is effective, applicable in IoT domains, and efficient in terms of accuracy, communication cost, energy consumption, and computational cost. 
It enables the implementation of secure federated learning on IoT devices.

The proposed xMK-CKKS based scheme is robust against honest-but curious participating devices, however, it may not withstand attacks of malicious participants that sabotage the learning process.  
For instance, instead of sharing correct gradient updates, malicious devices may send incorrect or random values or even values specifically designed to interfere and bias the federated learning process. 
A common mechanism to address this problem attempts to identify incorrect parameters to discard them by designing Byzantine robust aggregation rules. 
For further details, we refer to Median~\cite{yin2018byzantine} and Krum~\cite{blanchard2017machine}.

However, these defense mechanisms assume that the training data of each device is independent and identically distributed (iid), which is difficult to be guaranteed in actual IoT scenarios. 
More importantly, the server needs to obtain the model updates from each individual device, which again introduces a risk of privacy leakage to the server or other participants overhearing the communication. 
However, protecting the individual model updates via multi-key cryptographic routines, as the xMK-CKKS based federated learning scheme we have proposed, prevents that individual model updates can be analyzed for potential anomalies or suspicious patterns. 
Therefore, further research is needed to investigate how federated learning can not only resist Byzantine attacks, but also protect the privacy of the remotely participating devices.

\section*{acknowledgements}
This work is partially supported by China National Science Foundation under grant number 62072356, and the National Key Research and Development Program of Shaanxi under grant number 2019ZDLGY12-08.
We would like to acknowledge partial funding by the Academy of Finland in the project ABACUS (ICT 2023). The support provided by China Scholarship Council (CSC) during a visit of Jing Ma to Aalto University is also acknowledged (file No.201906960151).

\bibliography{literature}
\bibliographystyle{unsrt} 

\end{document}